\documentclass[openacc]{rstransa}
\usepackage[utf8]{inputenc}

\usepackage[a4paper, total={6.7in, 10in}]{geometry}

\usepackage{amsmath,amssymb,amsfonts,amsthm,graphicx,endfloat,endnotes,setspace,verbatim,geometry,times, helvet,courier,bm,url, dcolumn,booktabs,cleveref,braket,cleveref}
\usepackage[version=4]{mhchem}

\newcommand{\mutwo}{\frac{\hbar^2}{2\mu_2}}
\newcommand{\hot}{(\hat{\mathbf{r_2}}, \hat{\mathbf{r_1}})}
\newcommand{\PtSF}{{\hspace{0.2em}}^{\textrm{SF}}\Phi^{JM\epsilon}_{n'j'l'}(\hat{\mathbf{r_2}},\mathbf{r_1})}
\newcommand{\PtSFnop}{{\hspace{0.2em}}^{\textrm{SF}}\Phi^{JM\epsilon}_{njl}(\hat{\mathbf{r_2}},\mathbf{r_1})}
\newcommand{\PtBF}{{\hspace{0.2em}}^{\textrm{BF}}\Phi^{JM\epsilon}_{n'j'\bar{\Omega'}}(\hat{\mathbf{r_2}},\mathbf{r_1})}
\newcommand{\PtBFnop}{{\hspace{0.2em}}^{\textrm{BF}}\Phi^{JM\epsilon}_{nj\bar{\Omega}}(\hat{\mathbf{r_2}},\mathbf{r_1})}

\newcommand{\YtSF}{{\hspace{0.2em}}^{\textrm{SF}}\mathcal{Y}_{jl}^{JM\epsilon}(\hat{\mathbf{r_1}},\hat{\mathbf{r_2}})}
\newcommand{\YtSFp}{ {\hspace{0.2em}}^{\textrm{SF}}\mathcal{Y}_{j'l'}^{JM\epsilon}\hot }
\newcommand{\YtBF}{{\hspace{0.2em}}^{\textrm{BF}}\mathcal{Y}_{j\bar{\Omega}}^{JM\epsilon}\hot}
\newcommand{\YtBFp}{ {\hspace{0.2em}}^{\textrm{BF}}\mathcal{Y}_{j'\bar{\Omega'}}^{JM\epsilon}\hot}
\newcommand{\FBF}{{\hspace{0.2em}}^{\textrm{BF}}F^{JM\epsilon}_{s E,nj\bar{\Omega}}(r_2)}
\newcommand{\FBFp}{{\hspace{0.2em}}^{\textrm{BF}}F^{JM\epsilon}_{s E,n'j'\bar{\Omega'}}(r_2)}
\newcommand{\FSF}{{\hspace{0.2em}}^{\textrm{SF}}F_{njl}^{JM\epsilon}(r_2)}
\newcommand{\FSFp}{{\hspace{0.2em}}^{\textrm{SF}}F_{n'j'l'}^{JM\epsilon}(r_2)}

\newcommand{\Pn}{P_{l\bar{\Omega}}^{JM\epsilon:j}}

\newcommand{\Ppp}{ P_{l'\bar{\Omega}}^{JM\epsilon:j'}  }
\newcommand{\Pppp}{ P_{l'\bar{\Omega'}}^{JM\epsilon:j'}   }
\newcommand{\LHS}{0}

    \newcommand{\tj}[6]{ \begin{pmatrix}
       #1 & #2 & #3 \\
       #4 & #5 & #6 
    \end{pmatrix}}

\newcommand{\jA}{A}
\newcommand{\jB}{B}
\newcommand{\jC}{C}

\begin{document}
\title{General Mathematical Formulation of Scattering Processes in Atom-Diatomic Collisions in the RmatReact Methodology}
\author{Laura K. McKemmish,$^{1,2}$  Jonathan Tennyson$^{2}$}
\address{}
\subject{}
\corres{l.mckemmish@unsw.edu.au, j.tennyson@ucl.ac.uk}
\jname{rsta}
\Journal{Phil. Trans. R. Soc}

\begin{abstract}
Accurately modelling cold and ultracold reactive collisions occuring over deep potential wells, such as \ce{D+ + H2 -> H+ + HD}, requires the development of new theoretical and computational methodologies. One potentially useful framework is the R-matrix method adopted widely for electron-molecule collisions which has more recently been applied to non-reactive heavy particle collisions such as Ar-Ar. The existing treatment of non-reactive elastic and inelastic scattering needs to be substantially extended to enable modelling of reactive collisions: this is the subject of this paper. Herein, we develop the general mathematical formulation for non-reactive elastic and inelastic scattering, photo-association, photo-dissociation, charge exchange and reactive scattering using the R-matrix method. Of particular note is that the inner region, of central importance to calculable R-matrix methodologies, must be finite in all scattering coordinates rather than a single scattering coordinate as for non-reactive scattering. 
\end{abstract}



\maketitle

\section{Introduction}

The rapid development of techniques for producing cold and even ultracold
molecules over the last decade is now enabling the study of chemical reactions and
scattering at the quantum scattering limit with only a few partial
waves contributing to the incident channel. Moreover, the ability to
perform these experiments with nonthermal distributions comprising 
specific states enables the observation and even full control
of state-to-state collision rates in this regime. This is perhaps the
most elementary study possible of scattering and reaction dynamics \cite{14StHuYe.Rmat}.
These experiments are driving the development of new theory to address the
new physics encountered in these ultra-slow collisions. Reactions
involved charged species are of special interest, in part because of the extra
experimental control possible for charge particles \cite{17Willit}. 


\begin{table}
\caption{Processes that can be studied using the {\sc RmatReact code}, using H$_2$D$^+$ 
system as an example.}
\begin{tabular}{ll}
\toprule
Process & Example \\
\midrule
Elastic collisions &  D$^+$ + H$_2(v,J)$ $\rightarrow$ D$^+$ + H$_2(v,J)$ \\
Inelastic collisions &   D$^+$ + H$_2(v'',J'')$ $\rightarrow$ D$^+$ + H$_2(v',J')$\\
Photodissociation   &  H$_2$D$^+$ + $h\nu$ $\rightarrow$ HD + H$^+$ or  H$_2$ + D$^+$ \\
Photoassociation & H$_2$ + D$^+$  $\rightarrow$ H$_2$D$^+$ + $h\nu$\\
Charge Exchange & D$^+$ + H$_2$ $\rightarrow$ D + H$_2^+$\\
Reactive scattering & D$^+$ + H$_2$ $\rightarrow$ HD + H$^+$ \\
\bottomrule
\end{tabular}\label{processes}
\end{table}

Due to the strength of the \ce{H3+} system interaction, the reaction of H$^+$ with H$_2$  can be particularly expected to show quantum behaviour at ultralow collision energies (temperatures);
this has meant that the H$_3^+$ system has become a benchmark system for the study of ultracold
reactions \cite{08CaGoRo,11HoJoGo,13HoSc,14GoScHo,14RaMaHo,14GoHo,15LaJaAo.Rmat}. 
Processes of interest are described in \Cref{processes} using the \ce{H2D+} system as an example. 
Other near-dissociation properties of H$_3^+$ that merit study include the  near-infrared photodissociation spectrum  
which was extensively chararacterised by Carrington, McNab and co-workers 
\cite{CB82,CB84,cm89,cmw92,00KeKiMc.H3+}, but which remains poorly
understood \cite{jt257}. The long-range H$^+$ -- H$_2$ potential has been shown
to support diffuse states which were called asymptotic vibrational states \cite{jt358}.
These states have some similar characteristics to halo states found in diatomic systems
\cite{19OwSp} but unlike the diatomic systems are likely to be present in significant numbers
although the actual density and structure of these states remains to be determined.
Finally,
formation of H$_3^+$ by radiative association could be important
in diffuse environments such as the early Universe. There are low temperature
measurements of this rate \cite{13GePlZy.H3+} but no theory and no studies at very low temperatures.

Procedures based on the use of hyperspherical coordinates \cite{87PaPa.Rmat,89LaLe.Rmat,18Kendri}
have been developed to solve the close-coupling equations, and applied to treat ultra-low energy reactive scattering in  D$^+$+H$_2$
see Lara \emph{et. al.} \cite{15LaJaAo.Rmat} and references therein. 

RmatReact is promising new methodology that addresses this problem by solving an initial energy-independent problem that encapsulates
most of the complicated physics in an inner region. These inner region solutions are then used to describe the scattering-dependent problem in the simpler outer region. This approach can be employed study not just the reactive scattering
process but, in principle, all the processes in \Cref{processes}. 
The RmatReact methodology described here is a spiritual successor to the extremely successful electron-molecule collision codes such as UKRMol \cite{jt474,jt518}, and indeed our implementation reuses part of this code base. Both methodologies separate space into three regions based on the distance between the two scattering partners: the inner, outer and asymptotic region. The inner region contains the region with significant non-multipole, and indeed non-local, interaction between the two scattering species; the Schrodinger equation solved here does not consider the scattering energy. The outer region contains the region where the interaction between the two scattering species is significant but simple in form (usually a multipole expansions); the outer region equations depend on the scattering energy and are solved once for every different scattering energy under consideration using the inner region solutions and a simple 1D propagation process. The asymptotic region is defined as the region in which the interaction between the scattering species is much less than the scattering energy and thus negligible; at this point, we can calculate properties of the overall scattering interaction such as cross-sections.

The mathematical formulation of our RmatReact methodology for non-reactive single and multi-channel scattering was previously presented by Tennyson \emph{et. al.} \cite{jt643}. Two initial applications of  this new RmatReact calculable R-matrix method for heavy-particle scattering to atom-atom collisions over the Morse potential and Ar-Ar scattering were presented by Rivlin \emph{et. al.}  \cite{jt727,jt755}.

Here, we extend the mathematical formulation of the RmatReact methodology to triatomics and consider for the first time photo-association, photo-dissociation, charge exchange and reactive scattering. 

\section{Mathematical Formulation}

\subsection{Overview of general RmatReact methodology}
Our methodology models scattering of atoms or molecules A and B with center of mass separation distance $R_{AB}$ using the following steps: 
\begin{enumerate}
    \item Variationally solve the (3$N$-3)D Schrodinger equation of the joint system AB in a finite region at zero scattering energy (including Bloch terms to ensure the Hermicity of the Hamiltonian) where $N$ is the total number of nuclei in the system and $R_{AB}$ less than some box size $a_0$: this produces a discrete number, $Z$, of inner-region energies, $E_i$, and wavefunctions, $\psi_i$.
    \item Map the (3N-3)D problem onto a 1D Hamiltonian in $R_{AB}$, with a reduced potential, $U(R_{AB})$. Simultaneously, each inner-region wavefunctions, $\psi_i$, can be mapped onto outer region channels, $\phi_c$ producing the surface amplitudes, $\omega_{c,i}$.
    \item Using the energy-independent  solutions, for each scattering energy under consideration, construct the scattering-energy-dependent R-matrix (to be defined below) at the boundary, thereby using the first set of solutions as an effective basis for describing the desired second set.  
    \item For each scattering energy, propagate this R-matrix to an asymptotic distance, at which point scattering observables such as cross-section can be evaluated using simple formulae.
\end{enumerate}

The Schrodinger equation solved in Step (i) is generally different for each system size. The bound state nuclear motion problem has been extensively considered in the context of high resolution spectroscopy studies \cite{jt626}. Introducing a finite region boundary into the problem does modify the Schrodinger equation somewhat and considerably change the nature of the solution, particularly in modified boundary conditions at $R_{AB}=a_0$ and through discretisation of the solutions above dissociation. 

The Schrodinger equation solved in Step (iii) has the same general form for all system sizes (and indeed is the same as the equations used in R-matrix theory for electron-atom and electron-molecule collisions aside from a reduced mass factor); a single program can hence be used in this outer region, with only numerical considerations (e.g. step-size) changing between systems. However, the form of the reduced potential, $U(R_{AB})$, changes with system size and is often non-trivial, particularly if coordinate transformations are involved. 

\subsection{Developing the mathematical description of scattering processes using the RmatReact methodology}

\subsubsection{Establishing the problem}
Much of the mathematics and analysis in this paper does not rely on the system being triatomic, and none relies on the system being \ce{H3+} or its isotopologues, though this will be used extensively as a illustrative and useful example of the methodologies discussed. 

\paragraph{Non-reactive elastic and inelastic scattering}
A triatomic non-reactive scattering problem, e.g. \ce{H+ + H2 -> H+ + H2}, is most effectively solved using Jacobi coordinates, with $r_1$ as the diatomic bond distance, $r_2$ as the distance between the centre of mass of the diatomic and the scattering atom, and $\theta$ as the angle between these two vectors.

\paragraph{Photo-association and photo-dissociation}
Non-reactive scattering coordinates are most appropriate here. These processes are half-collision processes involving scattering energy equal to the  photon energy, $h\nu$, minus the difference in energy between the dissociation energy $D_0$ and the energy of the initial state, $E_0$, i.e. E = $h\nu - D_0 + E_0$. 

\paragraph{Charge Exchange}
Treating charge exchange requires at least two electronic potential energy surfaces, called X and A here for simplicity. It is simplest to represent both potentials in a single coordinate system, the same as that used for non-reactive scattering.

\paragraph{Reactive scattering}
Consider a reactive scattering system which includes scattering channels with products A+BC, B+AC and C+AB with coordinates \jA, \jB{} and \jC{} respectively. Jacobi coordinates can be defined for each scattering coordinate when considering the triatomic reactive scattering case. Simplification to just two channels (i.e. modelling just a single reaction) is straightforward, and the extension to more than three channels logical. 


The mathematics described here builds on that presented in Chapter 7 of Burke \cite{Burke2011}; this has been successfully used for the study
of the positron-atom and positronium-ion reactive collision problem \cite{93HiBu}.

\subsubsection{Nuclear Motion Schrodinger Equation: Solving the Inner Region Hamiltonian at energy-independent }

The first stage of the RmatReact methodology is to find the wavefunctions and energies of the combined system in a finite inner region with zero scattering-energy. As these solutions will formally form a complete basis set in this finite region, these solutions can be used as a basis to describe the solutions to the scattering-energy-dependent problem in this inner region, i.e. 
\begin{equation}
\label{eq:BASIS}
    \Psi(E) = \sum_i A_i(E) \psi_i
\end{equation}
where $E$ is the scattering energy, $\Psi(E)$ is the inner region solution to the scattering-energy-dependent Schrodinger equation, $i$ count the solutions to the energy-independent  inner region problem $\psi_i$ and $A_i$ are expansion coefficients. 

In traditional quantum chemistry treatments, the full Schrodinger equation can be simplified by ignoring translational wavefunction and separating electronic, vibrational and rotational wavefunction. The separation of the electronic component is an approximation, known as the Born-Oppenheimer approximation \cite{27BoOp}, while the separation identification of the vibrational and electronic components is not an approximation provided the Corolois term is included (as we do).  The electronic component is considered in electronic structure packages to produce potential energy curves. In nuclear motion packages when treating a single electronic state, the total wavefunction can be represented as a sum of products between rotational and vibrational wavefunctions, i.e. $\psi_\textrm{total} = \sum_i \psi_i^\textrm{rot} \psi_i^\textrm{vib}$ (coefficients of the summation are absorbed into the vibrational wavefunction typically). The rotational wavefunction is a function of Euler angles $\alpha$, $\beta$, $\gamma$ and is quantised in terms of $J$ (the total angular momentum of the triatomic system), $M$ (the projected total angular momentum of the system onto the space-fixed $z$ axis), and $\Omega$ (the projected total angular momentum of the system onto the body-fixed $z$ axis) and described using Wigner $D$-functions \cite{57RoFe}, $D_{M\Omega}^{J*}$. Note in the absence of an external field, $M$ does not affect the energy of a molecular system and can be dropped from consideratio. Using this ansatz, the full Schrodinger equation is simplified to a set of $J$-dependent $3N-6$ dimensional Schrodinger equations that are typically given in internal vibrational coordinates.  The total wavefunction thus becomes 
\begin{equation} 
\psi_\textrm{total} = \sum_{M\Omega J} \psi_{J}^{vib} D_{M\Omega}^{J*}(\alpha, \beta, \gamma). 
\end{equation}  
This basis set, or appropriately symmetrised versions of it,  are used in variational nuclear motion programs, such as DVR3D \cite{DVR3D} for triatomic systems, to yield the vibrational wavefunctions, $\psi_{J}^\textrm{vib}$. For triatomics, using Jacobi coordinates  gives $\psi_J^\textrm{vib}(r_1, r_2, \theta)$.

The solutions to the traditional problem are bound state normalisable wavefunctions. In describing scattering, however, we need to include non-bound solutions corresponding to energies above the dissociation energy. Therefore, we move from an infinite region to a finite region, i.e. form a finite inner region, thereby discretising the continuum solutions. We ultimately desire the wavefunction solutions to the scattering Schrodinger equation at a large number of specific low scattering energies. An effective basis set to describe this large number of solutions can be formed by solving the single problem at zero scattering energy, as long as some of these solutions have non-zero value at the boundary between the inner and outer region (the R-matrix boundary). This is the first task of any R-matrix approach. 

Defining the inner region is a key component of the RmatReact methodology; slightly different concerns are necessary for each type of scattering process. 

\paragraph{Non-reactive elastic and inelastic  scattering}
 We formulate the inner region by introducing a finite domain constraint in the scattering coordinate, i.e. $0 \le r_2 \le a_0$ (utilising Jacobi  coordinates), where $a_0$ is known as the R-matrix boundary and defined as the scattering coordinate beyond which the two scattering systems interact via multipoles only to within the desired error, i.e. the potential beyond $a_0$ can be reduced from a full $(3N-6)$D potential to an effective 1D potential in the scattering coordinate $r_2$. In the inner region, we need at least one basis function which has a non-zero value at the R-matrix boundary $r_2 = a_0$ in order to describe the scattering wavefunction. We have found Lobatto shape functions \cite{88MaWy.Rmat,93Manolopoulos.Rmat,16WeissteinLobQuad.Rmat}
 to be a suitable choice of basis functions for this coordinate \cite{jt725,jt755}.
 
\paragraph{Photo-association, photo-dissociation}
The inner region for photo-association and photo-dissociation will be defined  as for non-reactive scattering with the additional caveat that the treatment of these processes \cite{bt75}  becomes more complicated \cite{86Seaton} if the quantum state of the combined system has significant magnitude beyond the R-matrix boundary (this case will not be considered here). In photo-dissociation, one would expect the reactant state to be well-bound and this will thus generally not be an issue. This is more likely to arise in photo-association when the product system (e.g. \ce{H2D+}) may be weakly bound in asymptotic halo states whose wavefunction is extended. 

\paragraph{Charge-exchange reactions}
The inner region R-matrix boundary $a_0$ should be defined such that it contains all non-negligible coupling between the two electronic states.  

\paragraph{Reactive scattering}  We need to expand this definition from a inner region defined by a single restricted domain to one that is constrained in all relevant scattering coordinates, i.e. $0 \le r_2^\jA \le a_0^\jA$, $0 \le r_2^\jB \le a_0^\jB$ and $0 \le r_2^\jC \le a_0^\jC$. Again, the three $a_0$ values should be chosen such that the interaction between $A$, $B$ and $C$ can be modelled as a function of the scattering coordinate alone. To ensure the Hermiticity of the Schrodinger equation in this restricted region, we construct Bloch operators \cite{blo57} at each of these boundaries like 
\begin{equation}
\mathcal{L}^\jA = \frac{1}{2} \delta(r_2^\jA - a_0^\jA) \frac{d}{dr_2^\jA}
\end{equation}
We then want to find the solutions to the Schrodinger equation with these Bloch terms, i.e. find $Z$ solutions $E_i$, $\Psi_i$ to
\begin{equation}
    (T + V +  \mathcal{L}^\jA + \mathcal{L}^\jB + \mathcal{L}^\jC) \psi_i = E_i \psi_i 
\end{equation}
where $T$ is the kinetic energy operator and $V$ is the potential energy operator for the system of interest. 
This is most easily done by using existing nuclear motion programs, e.g. DVR3D for triatomic systems, in which the Hamiltonian is modified to incorporate the Bloch terms and the basis functions are modified for this finite region. These inner region solutions $i$ will include both bound states of the molecule and a finite number of discretised continuum states. The strongly bound states will have energies and wavefunctions indistinguishable from the infinite region problem, but more weakly bound states will be influenced by the finite region constraints and be modified. As the box size increases, the differences for these weakly bound states will be smaller. The number of solutions will formally be equal to the number of basis functions utilised; however, we will only need to consider solutions with energies close to the scattering energy (often just above the dissociation energy of the reactant channel).

The Bloch terms and finite region introduces substantial requirements for the basis set, which must now ideally consist of basis functions with finite domain in three non-orthogonal scattering coordinates $r_2^\jA$,  $r_2^\jB$ and  $r_2^\jC$, with tractable resulting integrals. Furthermore, for each scattering boundary, at least one basis function must have a non-zero value but zero derivative; the zero derivative boundary condition introduced by necessity by the Bloch operator is a non-trivial and not often understood condition that has become obvious in our considerations of wavefunctions for RmatReact but which has been obscured in electron-molecule collision problems due to their far inferior basis sets: this fact is demonstrated in a simple system in the Appendix.  These multiple boundary conditions are unusual and non-trivial constraints on the design of the basis set that are not yet fully understood, and will be the most challenging part of utilising the RmatReact methodology for reactive collisions. We will thus defer its consideration to \Cref{sec:basisset}. We should, however, be reassured by the fact that this type of approach has been successfully utilised in light particle reactive collisions, e.g. where positron-atom reactants react to positronium-ion products \cite{Burke2011,93HiBu}.   




\subsubsection{Scattering Theory: Describing the Outer Region using Channels and Reduced Radial Functions}
The close-coupling equations \cite{ad60}  also simplifies the full Schrodinger equation in order to progress; however, instead of separating based on rotational and vibrational wavefunction, the wavefunction in the scattering coordinate (the reduced radial function) is separated from the other components of the wavefunction, which are described as channel in standard close-coupling treatments.  

\paragraph{Non-reactive scattering} 
The wavefunction in terms of channels and the reduced radial function, $F$ as 
\begin{equation}
\label{eq:NRchannels}
    \psi_\textrm{total} = \sum_c^\textrm{channels} \frac{1}{r_2} F_c(r_2) \Phi_c
\end{equation}
where the definition of $F$ is by convention, $c$ goes over all channel functions and $\Phi_c$ are the channel functions with all necessary coordinates orthogonal to $r_2$.  The product of the quantum states of the two isolated systems and their relative angular motion is typically used define the channels. For example, consider the triatomic system \ce{A + BC}. The scattering coordinate is $r_2^\jA$, the diatomic vibrational wavefunction is $\chi_n(r_1)$, the rotational angular momentum of the diatomic is $j$ and the relative angular momentum between the atom and diatomic is $l$, with the channel defined by $nlj$ and having energy equal to the diatomic vibrational and rotational energy.  Mathematically, we can use our knowledge of  the energies and wavefunctions of the asymptotic reactant and products that define the channel based on solving the bound state $(N-1)$-dimensional problem for which there are well-developed program solutions in the nuclear motion community; in the case of atom-diatomic scattering, this means using codes such as {\sc level} \cite{level} or {\sc Duo} \cite{Duo}. 
The form \Cref{eq:NRchannels} can in principle be used to describe the wavefunction in any region of space. However, for small $r_2^G$, the number of channels will be high especially if the combined system is strongly bound (e.g. \ce{H3+}, \ce{H2O}), i.e. the potential energy surface is deep compared to the vibrational spacings. This is one key reason why the inner region problem is solved separately using traditional nuclear motion techniques in the full dimensional space rather than using the equations in this section. In the outer region, however, for non-reactive scattering problem, only a small number of channels are important for describing systems, especially for ultracold collsions where often only a single rotational levels is populated.

The Schrodinger equation to be solved in the outer region for non-reactive scattering is given by the close-coupling \cite{ad60} expansion:

\begin{align}
\label{Eq:ODSE}
\left(-\mutwo\frac{\partial^2}{\partial r_2^2}+\mutwo \frac{l(l+1)}{r_2^2}+ (e_{nj} - E) \right) F_c(r_2) =&  - \sum_{c'}  \hat{U}_{c,c'}(r_2) F_{c'} (r_2)
 \end{align}
where $\mutwo$ is the  reduced mass along the Jacobu \lq scattering' coordinate,  $E$ is the scattering energy, $e_{nj}$ is the energy of the $n,j$ state of the diatomic and $U_{c,c'}(r_2)$ is the reduced potential $\braket{\phi_c|\Delta V|\phi_c'}$ with $\Delta V$ equal to the difference between the total potential energy operator for the triatomic system and the diatomic potential energy operator; thus $U_{c,c'} \rightarrow 0$ as $r_2$ increases. The values of $a_0$ should be determined such that $U_{c,c'}(r_2)$ is well represented by a multipole expansion in $r_2$. 

\paragraph{Photo-association, photo-dissociation}
In photo-association and photo-dissociation, only the reactant and product respectively consist of separated species (e.g. \ce{H2} and \ce{D+}) that are described by channels; the combined system (e.g. \ce{H2D+}) will be described by the quantum numbers of the combined system.

\paragraph{Charge-exchange}
The full wavefunction in the outer region will be described by channels in both electronic states, $c^X$, and $c_A$, i.e. 
\begin{equation}
    \psi_\textrm{total} =
\sum_{c_X} \frac{1}{r_2} F_{c_X}(r_2) \Phi_{c_X} +  \sum_{c_A} \frac{1}{r_2} F_{c_A}(r_2) \Phi_{c_A}. 
\end{equation}
The $c^X$ and $c_A$ channels are two sets of uncoupled channels. We can extend \Cref{Eq:ODSE} to sum over all $c=\{c^X,c_A\}$ as long as $U_{c^X,c_A} << E$ for all $c^X, c_A$, i.e. the interaction between channels on the different electronic surfaces are negligible compared to the scattering energy .

\paragraph{Reactive scattering}
In this more complicated case, we need to be able to deal with multiple scattering coordinates (i.e. $r_2^\jA, r_2^\jB, r_2^\jC$), and thus channels in multiple coordinates, $\Phi_{c_A}$, $\Phi_{c_B}$ and $\Phi_{c_C}$, and multiple reduced radial functions $F_{c_A}(r_2^\jA)$, $F_{c_B}(r_2^\jB)$ and $F_{c_C}(r_2^\jC)$. For generic coordinates, we will use `G' as our notation. We can thus write the total wavefunction as 
\begin{equation}
\label{eq:wfnouter}
    \psi_\textrm{total} =
\sum_{c_A} \frac{1}{r_2^\jA} F_{c_A}(r_2^\jA) \Phi_{c_A} +  \sum_{c_B} \frac{1}{r_2^\jB} F_{c_B}(r_2^\jB) \Phi_{c_B} +  \sum_{c_C} \frac{1}{r_2^\jC} F_{c_C}(r_2^\jC) \Phi_{c_C}
\end{equation}
where the coordinates of each of the channels are orthogonal to their associated $r_2^\jC$ coordinate. 

The number of channels in the outer region is dependent on the difference in energy between the reactant and product; modelling more exothermic reactive scattering processes will necessitate a much larger number of channels with a consequent considerable increase in the calculation time. 

To find $F_{c_G}(r_2^G)$ beyond the R-matrix boundary $a_0^G$, we reduce the full-dimensional Schrodinger to three one-dimension scattering Schrodinger equation in each set of coordinates $G=A, B, C$ equal to the \Cref{Eq:ODSE}. Note that by construction of the multiple R-matrix boundaries, the reduced potential $U$ connecting channels that scatter in different coordinates should be negligible compared to the scattering energy. 

\subsubsection{Energy-dependent Scattering using RmatReact methodology}

\paragraph{Non-reactive scattering}
The mathematics here has been discussed by Tennyson \emph{et. al.} \cite{jt643} and is summarised here to make clear the differences in treatment necessary between non-reactive scattering (the simplest kind of process) and the other processes, particularly reactive scattering. 

We have discussed earlier the fact that we solve the inner region energy-independent  problem in order to provide an efficient basis for describing the non-zero-scattering energy problem at the boundary, i.e. \Cref{eq:BASIS}. 
Mathematically, this approach utilises resolution of the identity (i.e. $1 = \sum_i \ket{i}\bra{i}$) and provides a spectral representation of the Green's function to find
\begin{equation}
\label{eq:NRpsiGreens}
    \ket{\Psi(E)} = \sum_{i=1}^Z \ket{\psi_i} \frac{1}{E_i - E} \braket{\psi_i|\mathcal{L} | \Psi(E)}
\end{equation}
where $\mathcal{L}$ is the Bloch term given by $\mathcal{L} = \mutwo \delta(r_2-a_0) \frac{d}{dr_2}$,  $\Psi(E)$ is the scattering-energy-dependent wavefunction and $E$ is again the scattering energy. The summation runs over all inner region solutions, $i$, which is formally infinite but in practice finite due to the representation of the inner region solutions in a basis set of size $Z$. As this equation has the desired solution $\Psi(E)$ on both sides, this representation cannot be directly utilised in a computational solution.

Instead, mathematical transformations described in Burke \cite{Burke2011} and Tennyson \emph{et. al.}  \cite{jt643} yield an expression linking the reduced radial functions $F$ and the R-matrix $R$ at the R-matrix boundary $r_2=a_0$ as
\begin{equation}
    F_c(a_0) = \sum_c R_{cc'}(E) a_0 \frac{dF_c}{dr_2}\Bigg|_{r_2=a_0}
\end{equation}
with the R-matrix defined by
\begin{equation}
R_{cc'}(E) = \frac{1}{2a_0} \sum_{i=1}^{Z} \frac{\omega_{c,i} \omega_{c',i}}{E_i-E}
\end{equation}
where the summation $i$ is over all inner region solutions, $c$, $c'$ go over all channels and $\omega_{c,i}$ are the surface amplitudes defined by
\begin{equation}
    \omega_{c,i} = \Braket{\frac{\phi_{c}}{r_2}| \psi_i}^\prime_{r_2 = a_0}
\end{equation}
where the prime in the Braket notation indicates the integral is over all coordinates except $r_2$.

Note that the coordinate systems for the inner and outer region, and quantum numbers for the inner region solutions and outer region channels need to be carefully considered and can have a substantial effect on the calculation time and accuracy. In particular, we highlight that inner-region codes are generally based on body-fixed coordinates, whereas outer-region propagation codes will generally use space-fixed coordinates to allow identification of the asymptotic channels with the states of the fragmented systems. Thus, a frame transformation is required to convert between these; this is discussed in Appendix A (which also defines the below quantum numbers and notation). The key integral that must be calculated is the surface amplitude between inner region solution $i$ and outer region channel $njl$, denoted as  $\omega_{njl,i}^{JM\epsilon}(a_0)$, where the $JM\epsilon$ denote the quantum numbers for the combined triatomic system. As made clear by the notation, the outer region channel function is usually defined in space-fixed (SF) coordinates as $\phi_{njl}^{JM\epsilon}$. The inner region solution, $^{BF}\psi_i^{JM\epsilon}$, is defined in body-fixed coordinates such that it can be expanded as  $^{BF}\psi_i^{JM\epsilon} =\sum_{n'j'\bar{\Omega'}} F_{n'j'\bar{\Omega'}}^{JM\epsilon}(r_2) ^{BF}\phi_{n'j'\bar{\Omega'}}^{JM\epsilon}$. Thus the surface amplitude can be calculated using
\begin{eqnarray}
        \omega_{njl,i}^{JM\epsilon}(a_0)  &=&  \Braket{\frac{^{SF}\phi_{njl}^{JM\epsilon}}{r_2}| ^{BF}\psi_{i}^{JM\epsilon}}^\prime_{r_2 = a_0} \\
        &=& \sum_{\bar{\Omega}} P_{l\bar{\Omega}}^{JM\epsilon:j}\Braket{\frac{^{BF}\phi_{nj\bar{\Omega}}^{JM\epsilon}}{r_2}| \sum_{n'j'\bar{\Omega'}} F_{n'j'\bar{\Omega'}}^{JM\epsilon}(r_2) ^{BF}\phi_{n'j'\bar{\Omega'}}^{JM\epsilon}}^\prime_{r_2 = a_0} \\
        &=& \sum_{n'j'\bar{\Omega'}} \frac{1}{a_0}F_{n'j'\bar{\Omega'}}^{JM\epsilon}(a_0) \sum_{\bar{\Omega}} P_{l\bar{\Omega}}^{JM\epsilon:j}\Braket{^{BF}\phi_{nj\bar{\Omega}}^{JM\epsilon}|  ^{BF}\phi_{n'j'\bar{\Omega'}}^{JM\epsilon}}^\prime_{r_2 = a_0} \\
        &=& \sum_{n'j'\bar{\Omega'}} \frac{1}{a_0}F_{n'j'\bar{\Omega'}}^{JM\epsilon}(a_0) \sum_{\bar{\Omega}} P_{l\bar{\Omega}}^{JM\epsilon:j} \delta_{nn'} \delta_{jj'} \delta_{\bar{\Omega}\bar{\Omega'}} \\
        &=& \sum_{\bar{\Omega}} \frac{1}{a_0}F_{nj\bar{\Omega}}^{JM\epsilon}(a_0)  P_{l\bar{\Omega}}^{JM\epsilon:j'}   \\
\end{eqnarray}
where we use the fact that the channels in body-fixed coordinates are defined such that they are orthonormal, i.e.  $\Braket{^{BF}\phi_{n'j'\bar{\Omega}'}^{JM\epsilon}|  ^{BF}\phi_{nj\bar{\Omega}}^{JM\epsilon}}^\prime_{r_2 = a_0} = \delta_{nn'}\delta_{jj'}\delta_{\bar{\Omega}\bar{\Omega'}}$.

\paragraph{Photo-association and photo-dissociation}
To describe photo-association and photo-dissociation processes using the R-matrix approach, we need to follow the approach to atomic photo-ionisation developed by Burke and Taylor \cite{bt75} which links the initial wavefunction with the final wavefunction in the inner regions through the electric dipole.

In the case of photo-dissociation, let the initial wavefunction be $\Psi_0$, an eigenstate of the energy-independent  inner region, i.e. $\psi_i$ for some $i$; this will often be the ground or low-lying rovibrational states of the combined system (e.g. \ce{H3+}). 
In the inner region, the final wavefunction in the inner region,  $\Psi^\textrm{inner}(E)$, can be written as a sum of inner region wavefunctions $\psi_f$ where $f$ sums over all inner region states as in \Cref{eq:BASIS}, where the photon energy, $h\nu$, is equal to the scattering energy, $E$, plus the difference in energy between the dissociation energy $D_0$ and the energy of the initial state, $E_0$, i.e. $h\nu = E + D_0 - E_0$. 
Then the integral that characterises the strength of the photodissociation process as a function of $h\nu$ is
\begin{eqnarray}
    \mu^\textrm{trans-pd}(h\nu)  &=&  \Braket{\Psi_0| \mu | \Psi^\textrm{inner}(h\nu - D_0 + E_0)} \\
    &=&  \sum_f A_f(h\nu - D_0 + E_0)  \Braket{\psi_i| \mu |\psi_f}
\end{eqnarray}
Then, we need to have the dipole moment function for the inner region so that $\Braket{\psi_i| \mu |\psi_f}$ can be calculated over the full finite inner region, as well as $A_f$, i.e. the coefficients of expansion for the scattering-energy-dependent inner region wavefunction in terms of the energy-independent  wavefunction which are generated using non-reactive scattering with half-collision boundary conditions \cite{bt75}. 

This approach should be capable of providing a complete model for the challenging H$_3^+$ near-dissociation spectrum of Carrington and coworkers \cite{CB82,CB84,cm89,cmw92,00KeKiMc.H3+} including, for example, modeling the different resonance widths observed in  their spectra.

The case of photo-association is analogous except that the initial and final states are switched, i.e. 
\begin{equation}
    \mu^\textrm{trans-pa}(h\nu) = \sum_i A_i(h\nu - D_0 + E_0)  \Braket{\psi_f| \mu |\psi_i},
\end{equation}
where this expression is only valid if the final wavefunction $\psi_f$ can be assumed to have negligible extent beyond the R-matrix boundary $a_0$. 
Within an R-matrix formulation it is also possible to consider the contribution of dipole transitions arising from the outer region \cite{86Seaton}, but this is beyond the scope of this paper.

\paragraph{Charge-exchange}
\Cref{eq:NRpsiGreens} applies for the charge exchange process with the channel index $c$ going over both the $c_X$ and $c_A$ channels.

\paragraph{Reactive scattering}
Following the same logic as for the non-scattering case, we obtain 
\begin{equation}
\label{eq:psiGreens}
    \ket{\Psi(E)} = \sum_{i=1}^n \ket{\psi_i} \frac{1}{E_i - E} \braket{\psi_i|\mathcal{L}_A + \mathcal{L}_B + \mathcal{L}_C | \Psi(E)},
\end{equation}
with the key difference here being the need for multiple R-matrix boundaries and thus multiple Bloch terms. As we need to consider multiple boundaries and their associated coordinates, there is a substantially more complex form for the reduced radial functions $F$ and for the R-matrix, $R$. Following the methodology for a related derivation in Burke \cite{Burke2011}, we start from \Cref{eq:psiGreens}, project onto channel functions $\phi_{c_G}$ (different for each coordinate), evaluate on the boundary, and ultimately obtain (in analogy with Eq. (7.26)--(7.27) of Burke  \cite{Burke2011}), yielding
\begin{equation}
    F_{c_G}^G(a_0^G) = \sum_{c'_A} R_{c_G c'_A}(E) a_0^A \frac{dF_{c'_A}}{dr_2^A}\Bigg|_{r_2^A=a_0^A} + \sum_{c'_B} R_{c_G c'_B}(E) a_0^B \frac{dF_{c'_B}}{dr_2^B}\Bigg|_{r_2^B=a_0^B} + \sum_{c'_C} R_{c_G c'_C}(E) a_0^C \frac{dF_{c'_C}}{dr_2^C}\Bigg|_{r_2^C=a_0^C}.
\end{equation}
and the R-matrix found by
\begin{equation}
 R_{c_G^{} c'_{G'}}(E) = \frac{1}{2a_0^{G'}} \sum_{i=1}^{Z} \frac{\omega_{c_G, i}\omega_{c'_{G'}i}}{E_i-E}
\end{equation}
where ${c_G}$ runs over all channels associated with coordinate $G$, i.e. $\jA, \jB, \jC$ if \ce{A + BC, B + AC, C + AB} are all considered explicitly, and where $\omega_{c,i}^G$ is the surface amplitude  defined as
\begin{equation}
\label{eq:surfaceamp}
    \omega_{c_G,i} = \Braket{\frac{\phi_{c_G}}{r_2^G}| \psi_i}^\prime_{r_2^G = a_0^G}
\end{equation}
Thus, with these equations, the R-matrix at a boundary $a_0^{G'}$ can thus be obtained for any scattering energy if the inner region problem can be solved and if the surface amplitudes can be computed.

There are a number of important things to note about these expressions. 
First, the R-matrix is not symmetric with respect to its indices, i.e. the denominator is a function only of the second coordinate boundary, $a_0^{G'}$. 
Second, the summation in this definition is formally over all inner region solutions to the energy-independent  problem; however in practice  bound state solutions corresponding to states well below dissociation will have near zero amplitudes at all three boundaries (i.e. $\omega_{ci}^G \approx 0, \forall G$) and can be  excluded from the summation without errors. It should also be possible to significantly reduce the sum over inner-region solutions with positive scattering to within, say, two orders-of-magnitude of the scattering energy, $E$, using the so-called partitioned R-matrix approach \cite{bb02,jt332} which uses simple formulae based on perturbation theory to correct the R-matrix for the contributions due to  higher energy poles. This ability to trim solutions should provide further contribute to the computational efficiency of this method. Finally, the key required integrals that must be evaluated to utilise these expressions are given by \Cref{eq:surfaceamp}, in which an inner region solution to the energy-independent  problem $\psi_i$ is projected onto an outer region channel. How this is done must be considered
carefully as it is this point in the calculation that one can introduce either a coordinate change and/or a frame transformation. Thus for example, one might wish to  project solutions on outer region channels which have different non-orthogonal scattering coordinates. Issues arising from this and possible
choices of inner-region basis sets are discussed in \Cref{sec:basisset}. 

The issue of multiple coordinates to consider does substantially increase the difficulty of this problem. Note, for example, how the surface amplitudes for the inner region solutions need to be evaluated in all scattering coordinates, A, B, C, in order to evaluate $R$.

\subsubsection{Propagation from R-matrix boundaries, and Asymptotic Expansion}
There are standard procedures available for R-matrix propagation \cite{lw76,bbm82,mor84}  which have been
widely and successfully used for both light and heavy particle scattering. Our proposed
solution is to use the parallel fast asymptotic R-matrix (PFARM) code \cite{PFARM}. Therefore considerations of how to propagate the R-matrix from the boundary to asymptotic distances, and to use asymptotic expansion and calculate scattering observables, is largely a solved problem. However, we make some notes in this section. 

\paragraph{Non-reactive scattering}
The propagation algorithm to go from the R-matrix boundary to its asymptotic value, and the way in which these asymptotic R-matrix is used to calculate the K-matrix and other scattering observables, is discussed for the single channel case in \cite{jt725}. Extensions to the multi-channel case are in progress, based on the use of PFARM in the outer region. 

\paragraph{Photo-association and photo-dissociation}
Propagation proceeds in the same manner as for non-reactive scattering, except an explicit form for the wavefunction must be calculated (i.e. the $A_i$, $A_f$), and that half-boundary-conditions must be applied \cite{bt75}. 

\paragraph{Charge exchange}
Techniques for propagating the R-matrix from the R-matrix boundary to asymptotic regions for sets of uncoupled channels is developed for the case of two sets of channels in Appendix E6 of Burke \cite{Burke2011}, and can be adopted with minimal changes to this problem. The crucial thing here is that two different electronic states are assumed be non-interacting in the outer region; this assumption provides the criterion for choosing an appropriate R-matrix boundary. Note that the R-matrices associated with the channels associatd with the two different surfaces are not always zero, otherwise the rate of the charge exchange process would be zero. Therefore, the propagation of the entire global R-matrix containing the two set of uncoupled channels needs to be performed at one time \cite{Burke2011}. 

\begin{figure}
    \centering
    \caption{Form of the K matrix for two uncoupled sets of channels.}
    \includegraphics[width=0.5\textwidth]{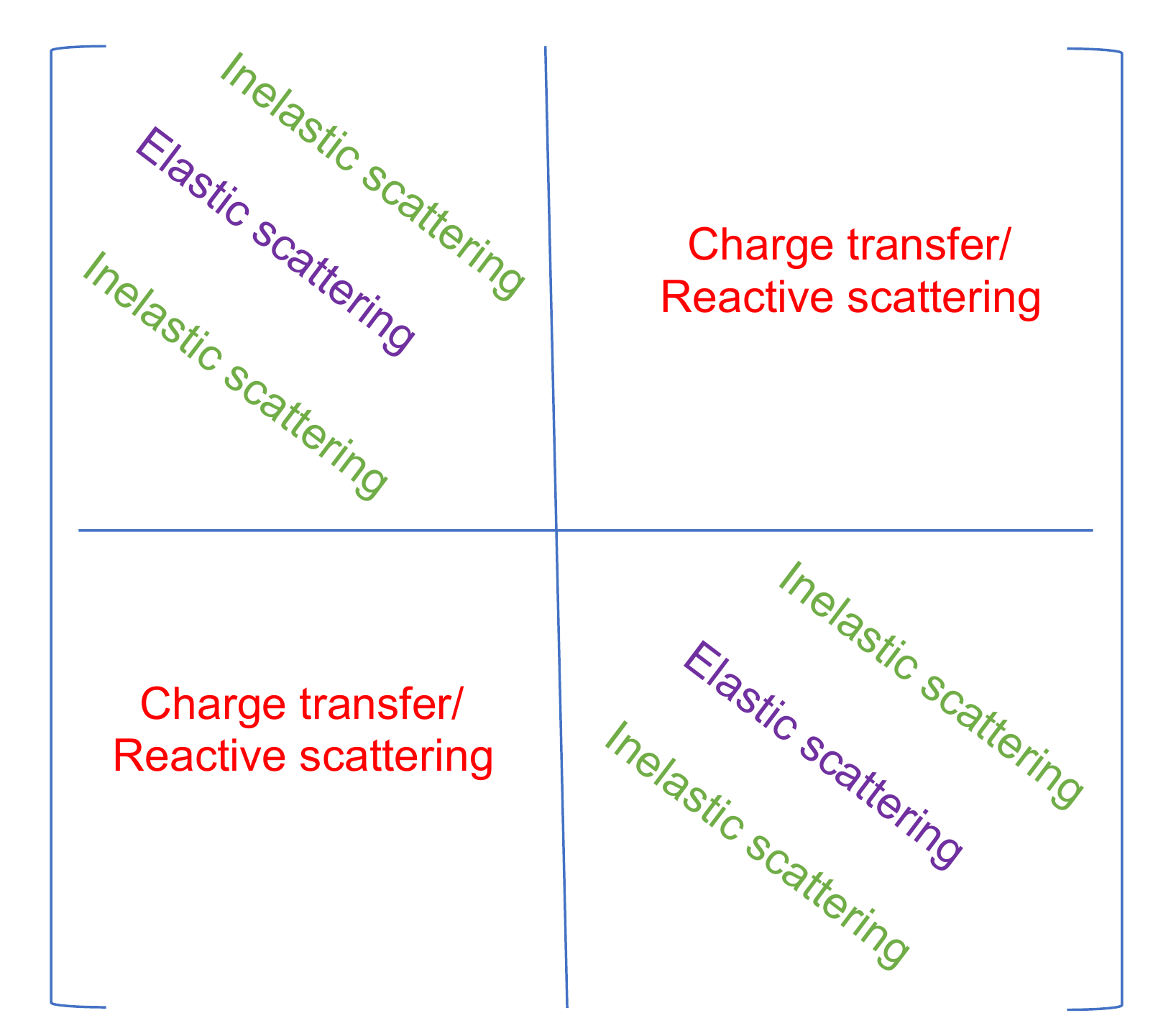}
    \label{fig:Kmat}
\end{figure}
The form of the K matrix expected from a charge exchange calculation is shown diagrammatically in \Cref{fig:Kmat}. Diagonal elements of this matrix represent elastic scattering processes, off-diagonal elements within a block represent inelastic processes and off-diagonal elements outside the two block diagonals represent charge transfer processes as the molecule moves from the X to A state.

\paragraph{Reactive scattering}
Techniques for propagating the R-matrix and reduced radial wavefunction from the R-matrix boundary to asymptotic regions for sets of uncoupled channels is developed for the case of two sets of channels in Burke \cite{Burke2011}; extensions to three sets of channels, if necessary for a particular problem, do not introduce fundamental changes to the approach here and will thus not be considered further in this paper. The mathematical  framework used to solve the outer region problem of reactively scattering atoms from positrons to form ions and positronium (considered by Burke) is highly analogous to the framework needed to describe reactive scattering of atoms and diatomics. A full computational implementation of this mathematical framework will be the subject of a future paper. 

The key required component for this outer region propagation is an expression of the reduced 1D potential energy curves, $U_{c,c'}$, defined in \Cref{Eq:ODSE}. A multipole expansion of the form \begin{equation}
    U_{c_G,c'_{G'}}(r_2) = \sum_{\lambda=1}^{\lambda^\textrm{max}} a_{c_G,c'_{G'},\lambda} (r_2^G)^{-\lambda-1} \delta_{G,G'}
\end{equation} 
is generally sufficient in the outer region, simplifying the propagation procedure. The value of the coefficients, $a_{c,c',\lambda}$, can be obtained through appropriate integration of the non-diatomic potential, $\Delta V$, over the channels in all coordinates other than the scattering coordinate of interest, $r_2^G$.  Thus for channels associated with  H$_2$ + H$^+$\ce{H3+}. Note that the reduced potential between channels in different scattering coordinates must be negligible (and set as zero) for this methodology to be implementable. 

The form of the K matrix for the full reactive scattering problem involving all three scattering coordinates will be a logical extension from \Cref{fig:Kmat}, with diagonal elements representing elastic scattering processes, off-diagonal in block elements representing inelastic processes and off-diagonal, off-block elements representing reactive scattering where the product and reactant are described using different Jacobi scattering coordinates.

\section{\label{sec:basisset}  Inner Region Coordinates and Basis Sets for Reactive Scattering}
For polyatomic systems the variational nuclear motion programs we plan to use to provide
solutions to the inner region problem, namely DVR3D \cite{DVR3D}, WAVR4 \cite{jt339} and TROVE \cite{trove}, all provide the choice of using a variety of different internal coordinates and
some control over the basis set employed. So far we assumed the use of Jacobi coordinates and have
not defined the basis sets will be used to compute the inner region energy-independent  solutions $\psi_i$. Basis sets consist of basis functions that are generally products of one-dimension basis functions in each coordinate of interest. In practice, the nuclear motion problems are generally solved on a discrete
variable representation (DVR) grid but transformation between polynomial basis functions and a DVR
is straightforward \cite{89BaLixx.method}, so it is sufficient at this stage to simply consider
basis functions.

For studies of the bound states of \ce{H3+} and its isotopologues including those up to dissociation, Jacobi coordinates have proven extremely successful \cite{jt65,jt100,jt236,jt512}, despite not representing the full symmetry of the problem. In describing these processes in the inner region problem, we use Laguerre polynomials (either Morse-like oscillators \cite{jt14} or spherical oscillator \cite{jt23}) for the \lq\lq diatomic\rq\rq $r_1$ Jacobi coordinates, (associated) Legendre polynomials for the $\theta$ coordinate and Lobatto shape functions  for the finite $r_2$ Jacobi coordinates. At the R-matrix boundary, we need at least one basis function to have a non-zero value (and formally the true solution has a zero-derivative boundary though this condition is less important); the need to satisfy this boundary condition and the finite domain in this coordinate is the main reason for utilising Lobatto shape functions.

For the reactive scattering problem a number of considerations need to be take into
account meaning that there a several possible options for internal coordinates. Explicitly, we consider: 
\begin{itemize}
\item A single set of Jacobi coordinates,
\item Multiple sets of  Jacobi coordinates,
\item Hyperspherical coordinates,
\item Radau coordinates.
\end{itemize}
Each of these coordinate choices leads naturally to a set of basis functions. 

When assessing our options, we want to consider a few factors. Ideally, we would like to consider the ingong and outgoing channels on an equal footing. Second (and most importantly), we need to be able to computationally effectively evaluate the overlap integrals and Hamiltonian matrix elements arising from the basis set and coordinates in a doubly- or triply-finite region (depending on how many scattering channels are energetically accessible) as well as evaluating the surface amplitude integrals (i.e. \Cref{eq:surfaceamp}).

\paragraph{Single Jacobi coordinate basis set}
For a triatomic reactive scattering code, it is potentially possible to use a traditional single set of Jacobi coordinate and define basis functions in this coordinate of the form \begin{equation}\{M_m (r_1) L_n (r_2) P_{j,k} (\theta)\}\end{equation} for a set of $m, n, j, k$, where $M_m$ are Laguerre polynomials, $L_n$ are Lobatto basis sets and $P_{j,k}$ are associated Legendre polynomials.

For this type of single-coordinate basis set runs into the following problems: 
\begin{enumerate}
\item The ingoing and outgoing channels are not treated equivalently;
\item The magnitude of the basis functions at the other R-matrix boundaries might not be sufficiently large to describe scattering properly in that coordinate; if the inner region basis functions cannot describe the wavefunction involved in scattering properly, then the RmatReact methodology will not be able to describe the scattering process properly;
    \item The zero-derivative boundary conditions will not be met in the other coordinate systems (this is a desirable but in practice not necessary condition);
    \item Accurately evaluating integrals with the finite boundary conditions imposed by the other scattering coordinate constraints is extremely complicated, especially as there will usually be no simple relationship between the coordinates.
\end{enumerate}

\paragraph{Multiple Jacobi coordinate basis sets}
To address some of the problems associated with using basis sets defined by a single set of Jacobi coordinates, we can introduce sets of basis functions associated with each Jacobi coordinate,  i.e. \begin{equation}\{M_{mA} (r_1^A) L_{nA} (r_2^A) P_{jA,kA} (\theta^A),M_{mB} (r_1^B) L_{nB} (r_2^B) P_{jB,kB} (\theta^B), M_{mC} (r_1^C) L_{nC} (r_2^C) P_{jC,kC} (\theta^C)\}
\end{equation} for a set of $mA, nA, jA, kA, mB, nB, jB, kB, mC, nC, jC, kC$.

This multiple-coordinate basis sets approach alleviates the first two problems for a single Jacobi coordinate basis set, but the problem of efficient computation of integrals for a non-orthogonal and over-complete basis will be necessary for the methodology to be of practical usefulness; this is not a general feature of variational nuclear motion programs.  
A related approach was successfully utilised by Day and Truhlar \cite{91DaTr}, who used  multiple Jacobi coordinate basis set to calculate the bound state energy levels of \ce{H3+}. This approach has the
additional advantage of restoring the full symmetry to the treatment of the \ce{H3+} problem.

The key difference between our problem and that solved by Day and Truhlar \cite{91DaTr} is that our problem has multiple boundary conditions. For example, one of the simpler integrals required is of the form
\begin{equation}
\int_0^{\pi} \int_0^\infty \int_0^{a_0^A}  F(r_1^A, r_2^A, \theta^A) H(a_0^B - r_2^B) dr_1^A dr_2^A d\theta^A 
\end{equation}
where $H(x)$ is the heavisidetheta function, i.e. $H(x)=1$ for $x>0$ and $H(x)=0$ for $x<0$. The best way to approach these integrals is to design the basis functions, $B$, such that $B(r_1^A, r_2^A, \theta^A) \ll E$ for $r_2^B>a_0^B$.  This is eminently feasible and will have the additional benefit of reducing linear dependency issues.

\paragraph{Hyperspherical coordinates}
Hyperspherical coordinates represent one way to avoid the need for multiple internal region coordinate sets and thus non-orthogonal basis sets. Hyperspherical coordinates  have already been used successfully for H$_+$ + H$_2$ reactive scattering \cite{14GoHo}. However, hyperspherical coordinates become increasingly inefficient at large scattering distances and thus applications to these sorts of problems have generally require transformation into Jacobi coordinates at some large scattering distance for efficient treatment of the problem \cite{15LaJaAo.Rmat}. Furthermore, the inner region nuclear motion calculations using hyperspherical coordinates are much less efficient than DVR codes based on orthogonal Jacobi and Radau coordinates \cite{jt566}.

\paragraph{Radau coordinates}
Problems where there are two reaction channels, such as \ce{D + OH -> DO + H}, can be efficiently treated using a single set of Radau coordinates (eg \cite{jt230,jt472,jt494}). In this case, one would use Lobatto shape functions (or Radau shape functions \cite{1880Radau.Rmat}) for both the $r_1$ and $r_2$ coordinates (with the $\theta$ coordinate basis functions being Legendre polynomials as usual). The surface amplitude integrals will thus require projection of the basis functions defined in Radau coordinates onto the channel basis functions that are naturally defined in the Jacobi $r_1$ and $\theta$ coordinates; such transformations are relatively easily performed in a DVR representation (see Appendix A of Tennyson {\it et al.} \cite{DVR3D}).


\section{Conclusions}
The RmatReact methodology is a new theoretical and computational methodology designed to treat ultracold heavy-particle scattering. The mathematical formalism borrows heavily from the highly successful calculable R-matrix methods that have been used extensively to treat electron-atom and electron-molecule collisions. However, there are crucial differences in the approach, particularly in the definition and solution for the inner region problem where the two scattering particles strongly interact. This paper extends for the first time the previously presented \cite{jt643} mathematical framework for treating atom-atom inelastic and elastic scattering to all atom-diatomic scattering processes relevant for the \ce{H3+} system: elastic and inelastic non-reactive scattering, photo-association and photo-dissociation, charge exchange and reactive scattering. The RmatReact methodology has the potential to revolutionise modelling of cold and ultracold heavy particle scattering by exploiting the inherent division of space into two regions: an inner region where the particle interactions are strong and should be treated in their full dimensionality with basis sets and calculation methods designed for molecular systems, i.e. nuclear motion methods, and the outer region where particle interaction is weak but must be considered to a very large interparticle distance due to the small collision energies.

Our detailed consideration of the mathematics required to describe all scattering processes using the new RmatReact methodology shows that the key difficulty is probably in the choice of coordinates and basis sets to describe reactive heavy-particle scattering problems like \ce{D+ + H2 -> H+ + HD}. This occurs because the inner region becomes finite as defined over every non-orthogonal scattering coordinate. The modifications to this inner region Schrodinger equation require that for each boundary at least one basis function must have a non-zero contribution, raising  challenges in defining an appropriate basis set and evaluating the resultant integrals with multiple boundary conditions. To describe reactions with only  two relevant reaction channels, e.g.  \ce{D + OH -> H + OD}, Radau coordinates seem to be the most logical path forward. For systems where three reaction channels are of interest, e.g. symmetric H$^+ + \ce{H2}$ collisions, use of multiple Jacobi coordinates or hyperspherical coordinates appear to offer the best prospects for success. 


Previous 
studies on the H$_3^+$ system for reactive problems, eg D$^+$ + H$_2$($v =
0, j = 0$), have been performed on potentials with accurate long-range behaviour but a
relatively poor representation of the H$_3^+$ well region. We have recently developed a global
H$_3^+$ ground state potential energy surface \cite{jt748} which joins the highly accurate
{\it ab initio} spectroscopic potential of Pavanello {\it et al.} \cite{jt526} with the 
correct treatment of the above dissociation and asymptotic regions due to Velilla  {\it et al.}
\cite{08VeLeAg.H3+} to provide an accurate global potential for the \ce{H3+} system.
This will be used for our future studies on this system.

\section*{Acknowledgement}
We thank Tom Rivlin and Eryn Spinlove for many helpful discussions during the course of this work.

\aucontribute{LKM carried out the main body of investigation and drafted the manuscript. JT conceived of the study and provided critical insight to develop and interpret the results. All authors read and approved the manuscript.}
\funding{This project has received funding from the European Union's Horizon 2020 research and
innovation programme under the Marie Sklodowska-Curie grant agreement No 701962.}

\bibliographystyle{rsta}


\section*{Appendix: Frame Transformation for Triatomic Systems}
\appendix{}

Channels are defined by three coordinates: the vibrational quantum number of the diatomic vibration $n$, the rotational quantum number of the diatomic $j$ and a relative angular momentum quantum number that is $l$ in space-fixed coordinates and $\bar{\Omega}$ in body-fixed coordinates. 
The inner region problem is generally solved in body-fixed coordinates with solutions labelled by $nj\bar{\Omega}$, while the outer region solutions utilise space-fixed coordinates $njl$. It is imperative to be able to convert between these representations. The frame transformation mathematics described here are adapted from Launay (1976) \cite{76Launay} to be appropriate to the present problem. 

In both coordinate systems, the $r_1$ coordinate is the diatomic vector and the $r_2$ coordinate is the vector between the centre-of-mass of the diatomic and the scattering atom. For simplicity, the angular coordinates associated with each of these vectors are often referred to collectively as $\bar{\mathbf{r_1}}$ and $\bar{\mathbf{r_2}}$; they are defined differently for each coordinate system as described below. 

The total angular momentum of the triatomic system is $J$, with a projection of $M$; each solution to the inner region problem, each channel, each full solution to the scattering problem and each reduced radial function solution to the scattering problem are labelled by these quantum numbers. 

\subsection*{Body-fixed coordinates}
In 2-angle embedding \cite{jt22}, the angular component of the wavefunction is given by
\begin{equation}
\label{eq:2angle}
\mathcal{Y}^{JM}_{j\bar{\Omega}}(\hat{\mathbf{r_2}},\hat{\mathbf{r_1}}) = \sqrt{\frac{(2J+1)}{4\pi}} Y_{j\bar{\Omega}} (\theta, \gamma) D_{M\Omega}^{J}(\alpha,\beta,0), 
\end{equation}
where $Y_{lm}(\theta, \phi)$ is a spherical oscillator function 
and $\bar{\Omega} = 0,1,.... \min(j,J)$ if $\epsilon=(-1)^J$ and $\bar{\Omega} = 1,.... \min(j,J)$ if $\epsilon=(-1)^{J+1}$, 
while in 3-angle embedding, it is instead given by
\begin{equation}
\label{eq:3angle}
\mathcal{Y}^{JM}_{j\bar{\Omega}}(\hat{\mathbf{r_2}},\hat{\mathbf{r_1}}) =  P_{j\bar{\Omega}}(\theta) D_{M\Omega}^{J*} (\alpha, \beta, \gamma).
\end{equation}
where $P$ is a Legendre polynomial.

The combined eigenfunctions of definite total parity are given by
\begin{equation}
\label{Eq:calY}
\mathcal{Y}_{j\bar{\Omega}}^{JM\epsilon}(\hat{\mathbf{r_2}},\hat{\mathbf{r_1}}) = \frac{1}{\sqrt{2(1+\delta_{\bar{\Omega}0})}}  \left( \mathcal{Y}_{j\bar{\Omega}}^{JM}(\hat{\mathbf{r_2}},\hat{\mathbf{r_1}}) + \epsilon(-1)^J \mathcal{Y}_{j(-\bar{\Omega})}^{JM}(\hat{\mathbf{r_2}},\hat{\mathbf{r_1}})  \right)
\end{equation}

The expression for the channels in body-fixed coordinates, labeled by $nj\bar{\Omega}$, are thus
\begin{equation}
\label{Eq:PhiBF}
{\hspace{0.2em}}^{\textrm{BF}}\Phi_{nj\bar{\Omega}}^{JM\epsilon}({\hat{\mathbf{r_2}}},\mathbf{r_1}) = \chi_n(r_1) \YtBF
\end{equation}

An inner region solution $\alpha$ wavefunction in terms of channel functions is given by
\begin{equation}
\Psi_{\alpha}^{JM\epsilon} = \sum_{nj\bar{\Omega}} \frac{1}{r_2} {\hspace{0.2em}}^{\textrm{BF}}F_{i,nj\bar{\Omega}}^{JM\epsilon} (r_2) {\hspace{0.2em}}^{\textrm{BF}}\Phi_{nj\bar{\Omega}}^{JM\epsilon}({\hat{\mathbf{r_2}}},\mathbf{r_1}) 
\end{equation}

A scattering-energy solution $s$ at scattering-energy $E$ in terms of body-fixed channel functions is given by
\begin{equation}
\label{Eq:PsiBF}
\Psi_{sE}^{JM\epsilon} (\mathbf{r_2}, \mathbf{r_1})= \sum_{nj\bar{\Omega}} \frac{1}{r_2} {\hspace{0.2em}}^{\textrm{BF}}F_{sE,nj\bar{\Omega}}^{JM\epsilon} (r_2) {\hspace{0.2em}}^{\textrm{BF}}\Phi_{nj\bar{\Omega}}^{JM\epsilon}({\hat{\mathbf{r_2}}},\mathbf{r_1}) 
\end{equation}

\subsection*{Space-fixed coordinates}
The combined basis function for the angular coordinates takes the components of the angular momentum functions for the diatomic ($j,m_j$) and the atom rel. to the diatomic ($l,m_l$) that contribute to state with quantum numbers $J, M$  \cite{76Launay}:
\begin{equation}
{}^{\textrm{SF}}\mathcal{Y}_{jl}^{JM\epsilon}\hot =  \sum_{m_j m_l} (-1)^{j-l+M}(2J+1)^{1/2} \tj{j}{l}{J}{m_j}{m_l}{-M} Y_{jm_j} (\hat{\mathbf{r_1}}) Y_{lm_l}(\hat{\mathbf{r_2}})  
\end{equation}

In space fixed coordinates, the channels are labelled by $njl$, and given by
\begin{equation}
\label{Eq:PhiSF}
{}^{\textrm{SF}}\Phi_{njl}^{JM\epsilon}(\hat{\mathbf{r_2}},\mathbf{r_1})  = \chi_{n}(r_1) \YtSF
\end{equation}

A scattering-energy solution $s$ at scattering-energy $E$ in terms of space-fixed channel functions is given by
\begin{equation}
\label{Eq:PsiSF}
\Psi_{sE}^{JM\epsilon} (\mathbf{r_2}, \mathbf{r_1}) = \sum_{njl} \frac{1}{r_2} {}^{\textrm{SF}}F_{sE,njl}^{JM\epsilon} (r_2) {}^{\textrm{SF}}\Phi_{njl}^{JM\epsilon}({\hat{\mathbf{r_2}}},\mathbf{r_1}) 
\end{equation}

\subsection*{Connecting body-fixed and space-fixed coordinates}

Define 
\begin{align}
\Pn = &\braket{ \YtSF | \YtBF}  
\end{align}

Then, since our functions are all real
\begin{align}
\Pn = &\braket{\YtBF | \YtSF} \\
= & (-1)^{J+\bar{\Omega}}\tj{j}{J}{l}{\bar{\Omega}}{-\bar{\Omega}}{0} \frac{\sqrt{2(2l+1)}}{\sqrt{(1+\delta_{\bar{\Omega}0})}}
\end{align}
%

Since    $\mathcal{Y}_{jl}^{\textrm{SF},JM\epsilon}\hot$ for all $l$ form a complete set of eigenfunctions in $\hat{\mathbf{r_1}},\hat{\mathbf{r_2}}$, we can use resolution of the identity to demonstrate that 
\begin{align}
\ket{\YtBF } &=
 \sum_{l} \ket{ {\hspace{0.2em}}^{\textrm{SF}}\mathcal{Y}_{jl}^{JM\epsilon}\hot}\braket{ {\hspace{0.2em}}^{\textrm{SF}}\mathcal{Y}_{jl}^{JM\epsilon}\hot | \YtBF} \notag  \\
& = \sum_{l} \Pn \ket{{\hspace{0.2em}}^{\textrm{SF}}\mathcal{Y}_{jl}^{JM\epsilon}\hot}
\label{Eq:YBF2SF}
\end{align}

Similarly,
\begin{equation}
\label{Eq:YSF2BF}
{}^{\textrm{SF}}\mathcal{Y}_{jl}^{JM\epsilon}\hot = \sum_{\bar{\Omega}} \Pn \YtBF
\end{equation}

\vspace{1em}

Equating the body-fixed and space-fixed representations of the full wavefunction, 
\begin{align}
\sum_{l} \ket{\FSF \YtSF}  =&\sum_{\bar{\Omega}} \ket{\FBF{\hspace{0.2em}}^{\textrm{BF}}\mathcal{Y}_{j\bar{\Omega}}^{JM\epsilon}(\hat{\mathbf{r_1}},\hat{\mathbf{r_2}})} \notag \\
=&\sum_{\bar{\Omega}} \sum_l \ket{\YtSF}\bra{\YtSF} \ket{\FBF\YtBF} \notag \\
 =&  \sum_l\ket{\YtSF}  \left(\sum_{\bar{\Omega}}\bra{\YtSF} \ket{\FBF\YtBF} \right) \notag \\
 =&  \sum_l \left(\sum_{\bar{\Omega}} \Pn  \ket{\FBF }  \right) \YtSF
\end{align}
Then, by comparing terms within the expression
\begin{align}
\label{Eq:FSF2BF}
\FSF = \sum_{\bar{\Omega}} \Pn \FBF
\end{align}
Similarly, 
\begin{align}
\label{Eq:FBF2SF}
\FBF = \sum_l \Pn \FSF
\end{align}

\end{document}